\documentclass[a4paper,11pt]{article}
\pdfoutput=1

\usepackage{jheppub}

\usepackage[abs]{overpic}

\usepackage{warpcol}

\usepackage{lineno}
\usepackage{graphicx}
\usepackage{dcolumn}
\usepackage{bm}
\usepackage{rotating}
\usepackage{epstopdf}
\usepackage{color}
\usepackage{verbatim} 
\usepackage{multirow}
\usepackage[abs]{overpic}
\usepackage{amsmath}
\usepackage{mathrsfs}
\usepackage{amssymb}
\usepackage{subfigure}
\usepackage{xspace}
\usepackage{float}

\newcommand{\PreserveBackslash}[1]{\let\temp=\\#1\let\\=\temp}
\newcolumntype{C}[1]{>{\PreserveBackslash\centering}p{#1}}
\newcolumntype{R}[1]{>{\PreserveBackslash\raggedleft}p{#1}}
\newcolumntype{L}[1]{>{\PreserveBackslash\raggedright}p{#1}}

\newcommand{\pp}{\pi^+\pi^-}

\newcommand{\EE}{e^+e^-}

\newcommand{\psip}{\psi(3686)}

\newcommand{\jpsi}{J/\psi}
\newcommand{\ks}{K_S^0}

\newcommand{\gev}{\mathrm{GeV}}

\newcommand{\mev}{\mathrm{MeV}}
\newcommand{\invfb}{\mathrm{fb}^{-1}}
\usepackage[T1]{fontenc}

\RequirePackage{lineno}
\usepackage{epstopdf}

\title{\boldmath Measurement of cross sections of $e^+e^-\to K^0_S K^0_S \psip$ from $\sqrt{s}=4.682$ to $4.951\,\gev$}

\collaborationImg{\includegraphics[width=.12\textwidth,origin=c,angle=90]{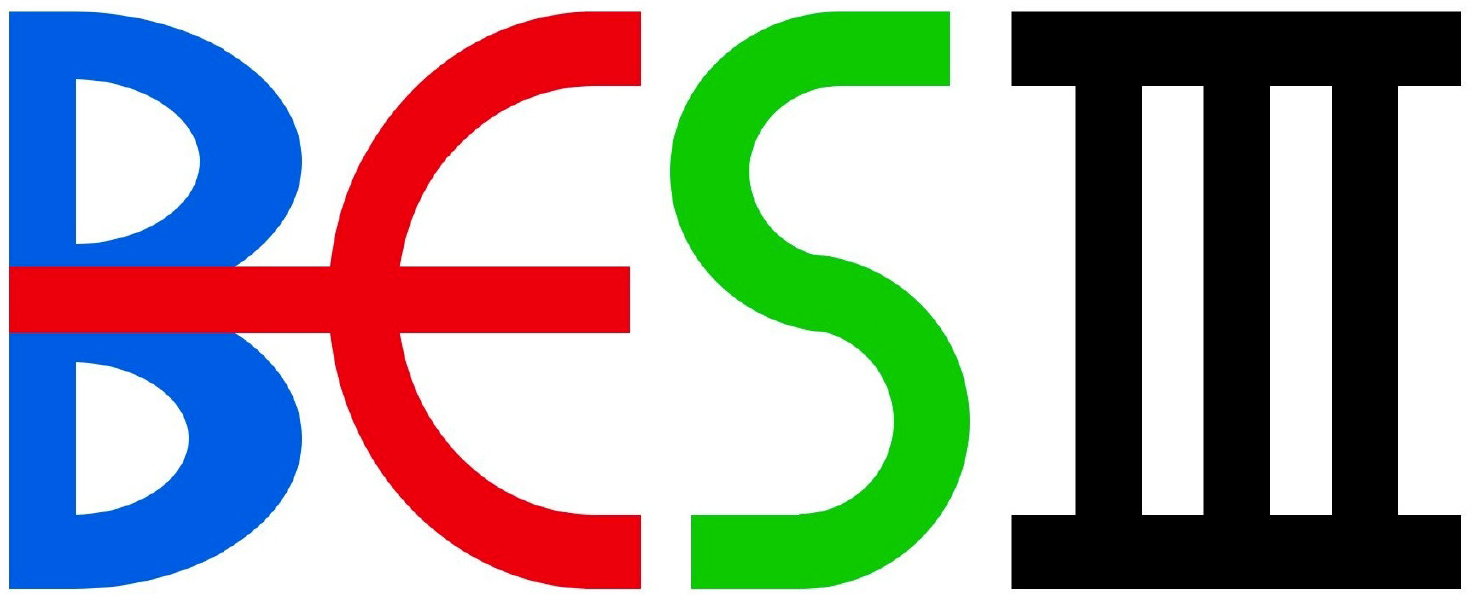}}

\collaboration{The BESIII Collaboration}


\emailAdd{besiii-publications@ihep.ac.cn}

\arxivnumber{1234.5678}

\abstract{
The process $e^+e^-\to K^0_S K^0_S \psi(3686)$ is studied by analyzing $e^+e^-$ collision data samples collected at eight center-of-mass energies ranging from 4.682 to 4.951 GeV with the BESIII detector operating at the BEPCII collider, corresponding to an integrated luminosity of $4.1\,\invfb$. Observation of the $e^+e^-\to K^0_S K^0_S \psi(3686)$ process is found for the first time with a statistical significance of $6.3\sigma$, and the cross sections at each center-of-mass energy are measured. The ratio of cross sections of $e^+e^-\to K_S^0 K_S^0 \psi(3686)$ relative to $e^+e^-\to K^+ K^- \psi(3686)$ is determined to be $\frac{\sigma(e^+e^-\to K_S^0 K_S^0 \psi(3686))}{\sigma(e^+e^-\to K^+ K^- \psi(3686))}=0.45 \pm 0.25$, which is consistent with the prediction based on isospin symmetry. The uncertainty includes both statistical and systematic contributions. Additionally, the $K_S^0\psi(3686)$ invariant mass distribution is found to be consistent with three-body phase space. The significance of a contribution beyond three-body phase space is only $0.8\sigma$.
} 

\begin{document}
\maketitle
\flushbottom

\section{Introduction}
Exotic hadrons, which are composed neither of a quark and anti-quark pair nor of three quarks, 
provide an unique opportunity to gain insight into the strong interaction.
Since the discovery of the $\chi_{c1}(3872)$ state by the Belle experiment in 2003~\cite{X3872}, 
numerous unexpected charmonium-like states, including the $Y(4260)$~\cite{Y4260-1, Y4260-2, Y4260-3}, $\psi(4660)$~\cite{Y4660-1, Y4660-2, Y4660-3, Y4660-4, Y4660-5}, $Z_c(3900)$~\cite{Zc3900-1,Zc3900-2} 
and $Z_{c}(4020)$~\cite{Zc4020}, as well as the $Z_{cs}(3985)$~\cite{Zc3985} and $Z_{cs}(4000)$~\cite{Zc4000}, 
have been observed by various experiments. The properties of these 
states do not match the predictions of conventional hadrons, and they are considered to be good 
candidates for exotic hadrons, 
such as hybrids~\cite{hybrids-1, hybrids-2, hybrids-3}, tetraquarks~\cite{tetraquarks-1}, meson molecules~\cite{molecules-1, molecules-2, molecules-3}, hadro-charmonium~\cite{hadrocharmonium-1, hadrocharmonium-2}, or a combination thereof~\cite{thereof}.

Several vector charmonium-like states have been reported based on a series of measurements of the cross sections of $\EE\to {\rm hadrons}$, including both 
hidden-charm and open-charm processes~\cite{Y4260-1, Y4260-3, BESIII:pipijpsi, BESIII:pipijpsi new, intro-BESIII-pipihc2, Y4660-4, BESIII:omegaChic0, BESIII:piDDstar, around 4.66-7, BESIII:etaJpsi, BESIII:etaJpsi new}. Three of these states, namely the $\psi(4660)$, $Y(4710)$, and $Y(4790)$, 
are located above $4.6\,\gev$. The $\psi(4660)$ resonance was first observed by the Belle Collaboration
in $\EE\to\pp\psip$ via initial-state-radiation (ISR)~\cite{Y4660-1}, was subsequently 
confirmed by the BaBar~\cite{Y4660-2} and BESIII~\cite{Y4660-4} Collaborations in the same process, and was also seen in $e^+e^-\to\pi^+\pi^-\psi_{2}(3823)$~\cite{Y4660-5} by the BESIII Collaboration.
Recently, BESIII measured the cross sections of 
$e^+e^-\to K\bar{K} J/\psi$~\cite{kkjpsi-2, intro-BESIII-kkjpsi, kkjpsi-4, intro-BESIII-ksksjpsi}, leading to the discovery of the $Y(4710)$. Additionally, a structure around $4.79\,\gev$,
referred to as the $Y(4790)$, was found to be essential to explain the cross section line-shape of 
$e^+e^-\to D^{*+}_{s}D^{*-}_{s}$~\cite{BESIII:dsds}. Potential models predict both the 5S and 4D charmonium states to be in this mass region~\cite{1-1, 1-2, 1-3, 1-4, 1-5}. Since the $Y(4710)$ and $Y(4790)$ are 
observed in final states containing $s\bar{s}c\bar{c}$ quark compositions, it is desirable to explore processes with similar quark compositions in the final state to uncover 
the nature of these vector structures.   

In addition to the vector states, the BESIII experiment also observed the $Z_{cs}(3985)^+$ state, the strange partner of 
the $Z_c$ states, in the process $e^+e^-\to K^{\pm} Z_{cs}(3985)^{\mp}, Z_{cs}(3985)^{\pm}$ $\to (\bar{D}^0 D^{*\pm}_{s}+\bar{D}^{*0} D^{\mp}_{s})$~\cite{Zc3985}.
Evidence of its neutral partner, the $Z_{cs}(3985)^0$, was also detected in the corresponding neutral process~\cite{ZCS(3985)0}.
Meanwhile, the LHCb experiment observed two tetraquark candidates, the $Z_{cs}(4000)^{+}$ and $Z_{cs}(4220)^{+}$, 
decaying into $K^{+} J/\psi$ in an amplitude analysis of $B^{+}\to K^{+}\phi J/\psi$~\cite{Zc4000}. 
There have been ongoing debates regarding whether the $Z_{cs}(3985)^+$ and $Z_{cs}(4000)^{+}$ are the same state, as 
they exhibit comparable masses but significantly different widths~\cite{ZCS(3900)-3, Zcs-1,Zcs-2,Zcs-3,Zcs-4,Zcs-5,Zcs-6,Zcs-7,Zcs-8}.
In the search for $Z_{cs}(3985)^{\pm}/Z_{cs}(4000)^{\pm}\to K^{\pm}\jpsi$ in $\EE\to K^+K^-\jpsi$ by the BESIII experiment, no significant signal for the $Z_{cs}^{+}$ was found~\cite{BESIII-kkpsip}. 

As an extension of the $\EE\to K^+K^-\jpsi$ analysis, the $\EE\to K^+K^-\psip$ process has also been investigated at BESIII
using data samples taken at center-of-mass (c.m.\/) energies $\sqrt{s}$ ranging from 4.699 to 4.951$\,\gev$~\cite{BESIII-kkpsip}. The cross section line-shape shows an enhancement around $4.843\,\gev$, while no $Z_{cs}^{\pm}$ signal is seen.

In this analysis, the neutral process $e^+e^-\to K^0_S K^0_S \psip$ is studied using data samples collected at eight c.m.\ energies from $4.682$ to $4.951\,\gev$ in 2020 and 2021, corresponding to a total integrated luminosity of $4.1\,\invfb$. The c.m.\ energies of the data samples are determined by selecting $\EE\to\Lambda^+_c\bar{\Lambda}^-_c$ events~\cite{cms-lumi-round1314},
and the luminosities are measured using large-angle Bhabha events~\cite{cms-lumi-round1314}.

\section{BESIII detector and data samples}

The BESIII detector~\cite{bes} records symmetric $e^+e^-$ collisions provided by 
the BEPCII storage ring~\cite{Yu:IPAC2016-TUYA01}, which operates with a peak luminosity 
of $1.1\times10^{33}$\,cm$^{-2}$s$^{-1}$ in the $\sqrt{s}$ range from $1.85$ to
$4.95$\,GeV~\cite{Ablikim:2019hff}. The cylindrical core of the BESIII detector covers $93\%$
of the full solid angle and consists of a helium-based multilayer drift chamber (MDC), 
a plastic scintillator time-of-flight system (TOF), and a CsI(Tl) electromagnetic 
calorimeter (EMC), which are all enclosed in a superconducting solenoidal magnet 
providing a $1.0$\,T magnetic field. The solenoid is supported by an octagonal
flux-return yoke with resistive plate counter muon identification modules interleaved 
with steel. The charged-particle momentum resolution at $1\,\gev/c$ is $0.5\%$,
and the ${\rm d}E/{\rm d}x$ resolution is $6\%$ for electrons from Bhabha scattering. The EMC measures photon energies with a resolution of $2.5\%$ ($5\%$) 
at $1\,\gev$ in the barrel (end-cap) region. The time resolution in the TOF barrel
region is $68$\,ps, while that in the end-cap region is $110$\,ps. The end-cap TOF
system was upgraded in 2015 using multi-gap resistive plate chamber technology, providing a time resolution of 60\,ps, which benefits all of the data used in this analysis~\cite{etofa,etofb,etofc}.

Monte Carlo (MC) samples are used to determine the detection efficiencies and to 
estimate the background contributions. MC samples are produced with a 
{\sc geant4}-based~\cite{geant4} package, which includes the geometric description 
of the BESIII detector and the detector response. The simulation models the beam 
energy spread and initial state radiation in the $e^+e^-$ annihilations with 
the generator {\sc kkmc}~\cite{ref:kkmca,ref:kkmcb}. All particle decays are simulated by {\sc evtgen}~\cite{ref:evtgen} with branching fractions either taken from the 
Particle Data Group (PDG)~\cite{PDG2018}, when available, or otherwise estimated with 
{\sc lundcharm}~\cite{ref:lundcharm}. Final state radiation from charged final state 
particles is incorporated using {\sc photos}~\cite{photos}. 
The inclusive MC samples include the production of open charm processes, the ISR 
production of vector charmonium(-like) states, and the continuum process incorporated 
in {\sc kkmc}~\cite{ref:kkmca,ref:kkmcb}.

In the MC simulation of the signal process, the $\psip$ decays to a $J/\psi$ and other particles, 
including 
$\psip\to J/\psi~+~( \pi^+\pi^-$, $\pi^0\pi^0$, $\eta$, $\pi^0$, $\gamma\gamma)$, 
where $\gamma\gamma$ refers to the process $\psip\to\gamma\chi_{cJ},~\chi_{cJ}\to\gamma\jpsi$. 
The numbers of events generated for the different decay modes of the $\psip$ are normalized according to the corresponding branching fractions.

\section{Event selection and background analysis}
To identify candidate events of interest, each of the two $\ks$ is reconstructed using $\ks \to \pp$, the $\psip$ is reconstructed using $\psip\to\jpsi+X$, and the $J/\psi$ using $J/\psi\to l^+l^-$ ($l=e,\mu$) decays. The particles accompanying the $\jpsi$ in the $\psip$ decays, denoted by $X$, are not reconstructed in order to increase statistics. The following criteria are applied to select candidate events for the final state $\pi^+\pi^-\pi^+\pi^-l^+l^-X$.

A charged track must have a distance of closest approach to the interaction point (IP) less than 10 cm along the $z$ axis ($|V_{z}|<10.0$\,cm) and less than 1 cm in the transverse plane ($|V_{xy}|<1$\,cm), and a polar angle ($\theta$) range of $|\cos\theta|<0.93$. The $z$ axis is the symmetry axis of the MDC, and $\theta$ is defined with respect to the $z$ axis. In addition, the momentum of a lepton candidate ($e$ or $\mu$) is required to be greater than 1.2\,$\gev/c$. At least two leptons with opposite charge are required to be present in the candidate event.

\begin{figure*}[htbp]
\begin{center}
\includegraphics[width=0.33\textwidth]{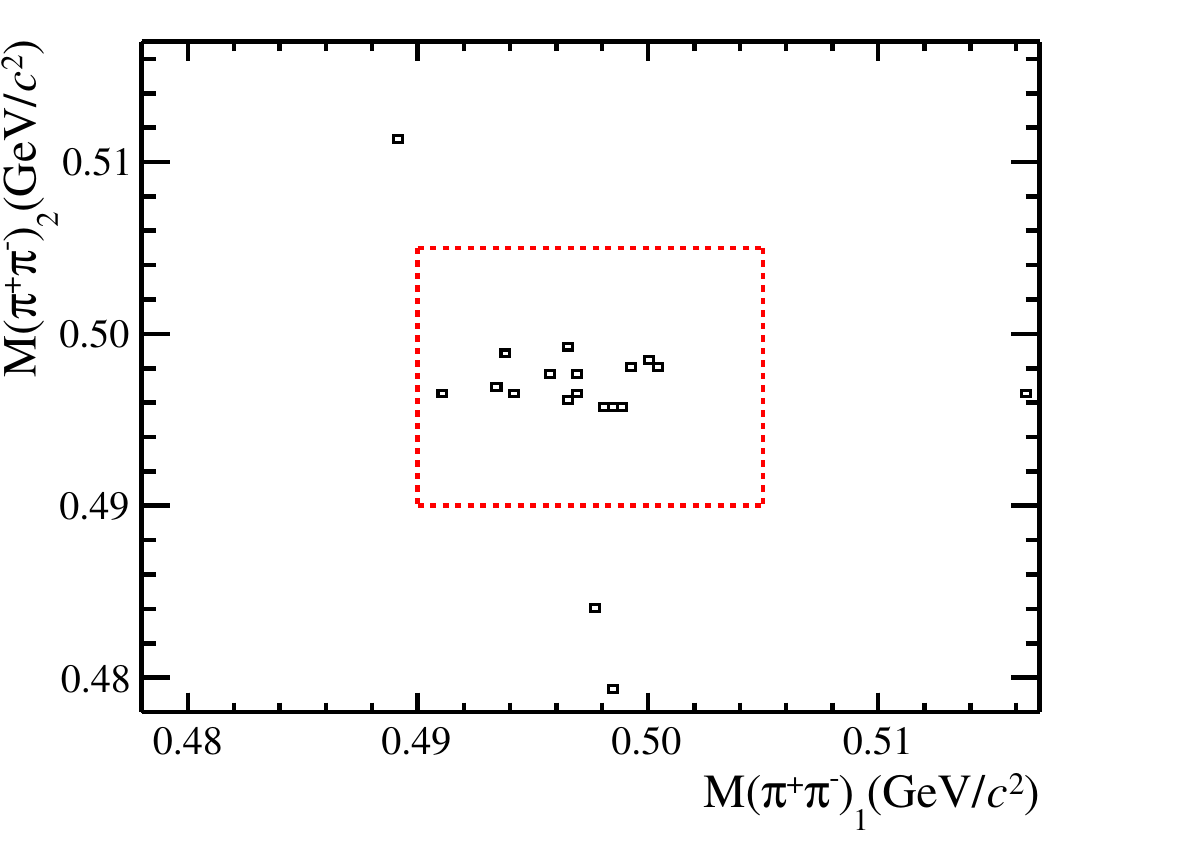}
\put(-80,-8){(a)}
\includegraphics[width=0.33\textwidth]{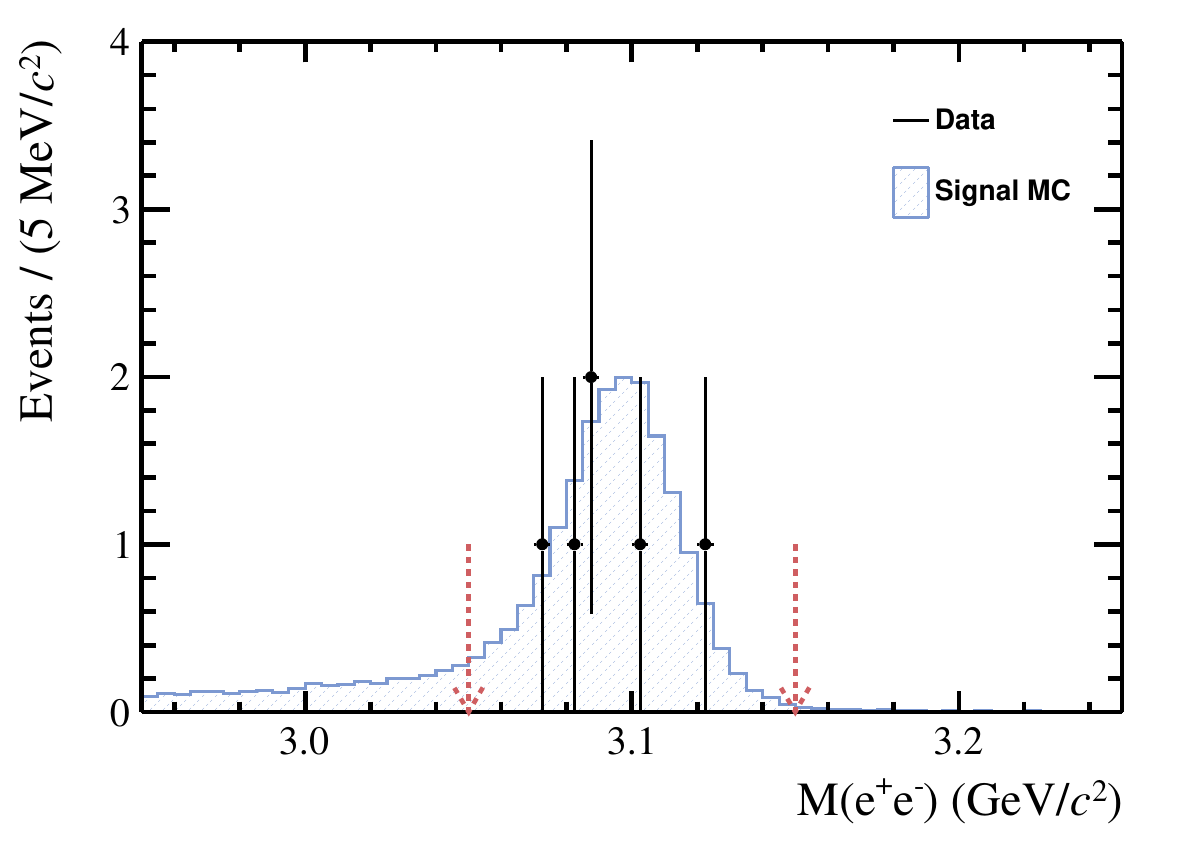}
\put(-80,-8){(b)}
\includegraphics[width=0.33\textwidth]{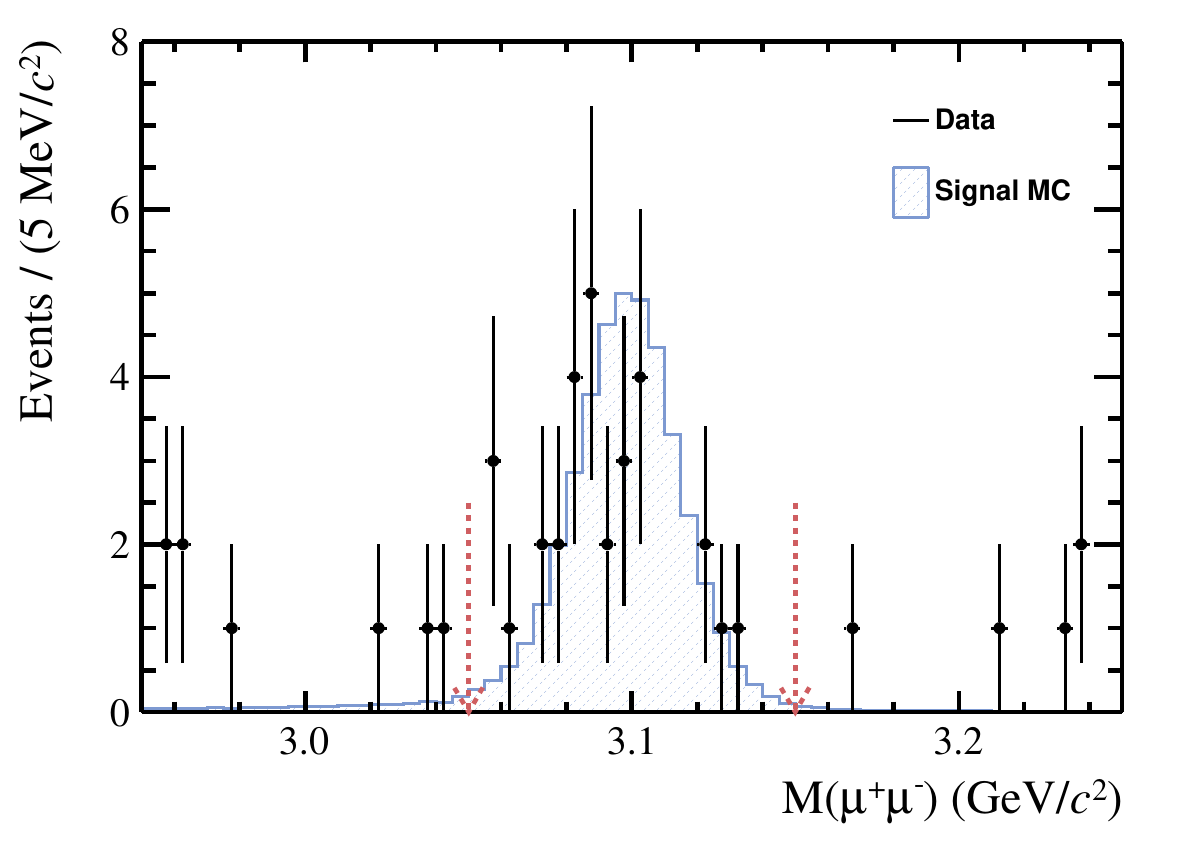}
\put(-80,-8){(c)}
\end{center}
\caption{(a) The scatter plot of $M(\pi^+\pi^-)_{1}$ versus $M(\pi^+\pi^-)_{2}$ from the sum of all data, where the red square denotes the $\ks$ signal region. (b) The $M(e^+e^-)$ and (c) the $M(\mu^+\mu^-)$ distribution obtained from the sum of all data, where the dots with error bars are data, the blue histograms are the sum of all signal MC samples at different energy points normalized according to the integrated luminosity, and the red dashed lines indicate the $\jpsi$ mass window.
}
\label{fig: ksks}
\end{figure*}

To select charged pions from $K^0_S$ decays, the $|V_{z}|$ and $|V_{xy}|$ requirements are not applied. Assuming they are pions, a vertex fit is performed for each pair of charged tracks. To reject random $\pi^+\pi^-$ combinations, a secondary-vertex fit algorithm is employed to impose a kinematic constraint between the production and decay vertices~\cite{ref:second vertexfit}. The $K^0_S$ mass window is set to be (0.490,~0.505)\,$\gev/c^2$. For each signal event candidate, there must be at least one $K^0_S K^0_S$ pair without overlapping pions. If there are two or more $K^0_S K^0_S$ pair candidates in the event, all combinations are kept for further study. The scatter plot of $M(\pi^+\pi^-)_{1}$ versus $M(\pi^+\pi^-)_{2}$ of the selected $\ks\ks$ pairs from all data samples combined is shown in figure~\ref{fig: ksks}(a).

A lepton candidate with more than 0.8\,$\gev$ of energy deposited in the calorimeter is assigned to be an electron, and a candidate depositing less than 0.5\,$\gev$ is assigned to be a muon~\cite{BESIII-kkpsip}. To improve the efficiency, this requirement is also applied to just one of the two leptons. If both tracks satisfy this requirement, or if one track does not have EMC information and the other track satisfies this requirement, they are both retained. The invariant mass of the lepton pair $M(l^+l^-)$ is used to identify the $J/\psi$, within a mass window of $(3.05,~3.15)\,\gev/c^2$, according to the mass resolution obtained from the simulation. The $M(l^+l^-)$ distributions from all data samples are shown in figures~\ref{fig: ksks}(b) and \ref{fig: ksks}(c).

\begin{table}[htbp]
\caption{
The numerical results at each c.m.\ energy, where the c.m.\ energies~($\sqrt{s}$) are rounded to the nearest $\mev$, $N_{\rm sdb}$ is the number of events in the $\psi(3686)$ sideband region, $N_{\rm obs}$ is the number of events in the $\psi(3686)$ signal region, $N_{\rm sig}$ is the calculated signal yield in the $\psi(3686)$ signal region, $\epsilon$ is the detection efficiency for corresponding lepton pair, ($1+\delta$) is the ISR correction factor, $\delta^{\rm {VP}}$ is the vacuum polarization factor, $\sigma_{\rm{Born}}$ is the Born cross section, $\sigma_{\rm{Born}}^{\rm{C.I.}}$ is the C.I. at the 90\% confidence level after taking into account  systematic uncertainty, and $\mathcal S$ is the statistical significance.
}
\label{tab:result}
\begin{center}
\scalebox{0.8}{
\begin{tabular}{cccccccccccc}
\hline
$\sqrt{s}$  &$N_{\rm sdb}$ &$N_{\rm obs}$ &$N_{\rm sig}$ & $N_{\rm sig }^{\rm{C.I.}}$ 
&$\epsilon_{ee}$  &$\epsilon_{\mu\mu}$  & ($1+\delta$)  
&$\delta^{\rm {VP}}$
& $\sigma_{\rm{Born}}$  &$\sigma_{\rm{Born}}^{\rm{C.I.}}$  
& $\mathcal S$ 
\\
(GeV)  & & & &  &(\%) &(\%) &  & & ($\rm{pb}$) & ($\rm{pb}$)  & ($\sigma$) 
\\
\hline
4.682 &1 &0 &$-0.5^{+1.1}_{-0.4}$  &(0.0, 2.0) &18.3 &26.3  &0.97 &1.05 &$-0.0^{+0.1}_{-0.0}\pm0.00$   &(0.0, 0.1) &-\\
4.699 &0 &0 &0  &(0.0, 2.0)                             &20.8 &29.8  &1.19 &1.05 &$\phantom{-}0.0^{+0.2}_{-0.0}\pm0.00$  &(0.0, 0.3) &-\\
4.740 &0 &2 &$\phantom{-}2.0^{+2.6}_{-1.3}$ &(0.5, 5.3) &21.9 &31.3  &0.86 &1.05 &$\phantom{-}1.4^{+1.9}_{-0.9}\pm0.10$ &(0.4, 3.8)  &1.2$\sigma$ \\
4.750 &1 &4 &$\phantom{-}3.5^{+3.4}_{-2.0}$ &(0.0, 8.0) &22.0 &31.0  &0.71 &1.05 &$\phantom{-}1.4^{+1.3}_{-0.8}\pm0.09$  &(0.0, 3.1) &1.7$\sigma$ \\
4.781 &2 &2 &$\phantom{-}1.0^{+2.9}_{-1.4}$ &(0.0, 4.8) &21.5 &30.6  &0.86 &1.06 &$\phantom{-}0.2^{+0.7}_{-0.4}\pm0.01$ &(0.0, 1.1) &0.2$\sigma$\\
4.843 &2 &6 &$\phantom{-}5.0^{+3.8}_{-2.5}$ &(0.6, 10.5) &16.5 &23.8  &0.89 &1.06 &$\phantom{-}1.4^{+1.1}_{-0.7}\pm0.09$  & (0.2, 3.0) &2.1$\sigma$ \\
4.918 &3 &1 &$-0.5^{+2.7}_{-1.2}$  &(0.0, 3.7) &15.3 &21.9  &1.05 &1.06 &$-0.3^{+1.8}_{-0.8}\pm0.02$ &(0.0, 2.5) &- \\
4.951 &0 &0 &0  &(0.0, 2.0)                              &12.4 &17.9  &1.20 &1.06 &$\phantom{-}0.0^{+1.0}_{-0.0}\pm0.00$ &(0.0, 1.8) &- \\
\hline
\end{tabular}}
\end{center}
\end{table}

\begin{figure*}[htbp]
\begin{center}
\includegraphics[width=0.45\textwidth]{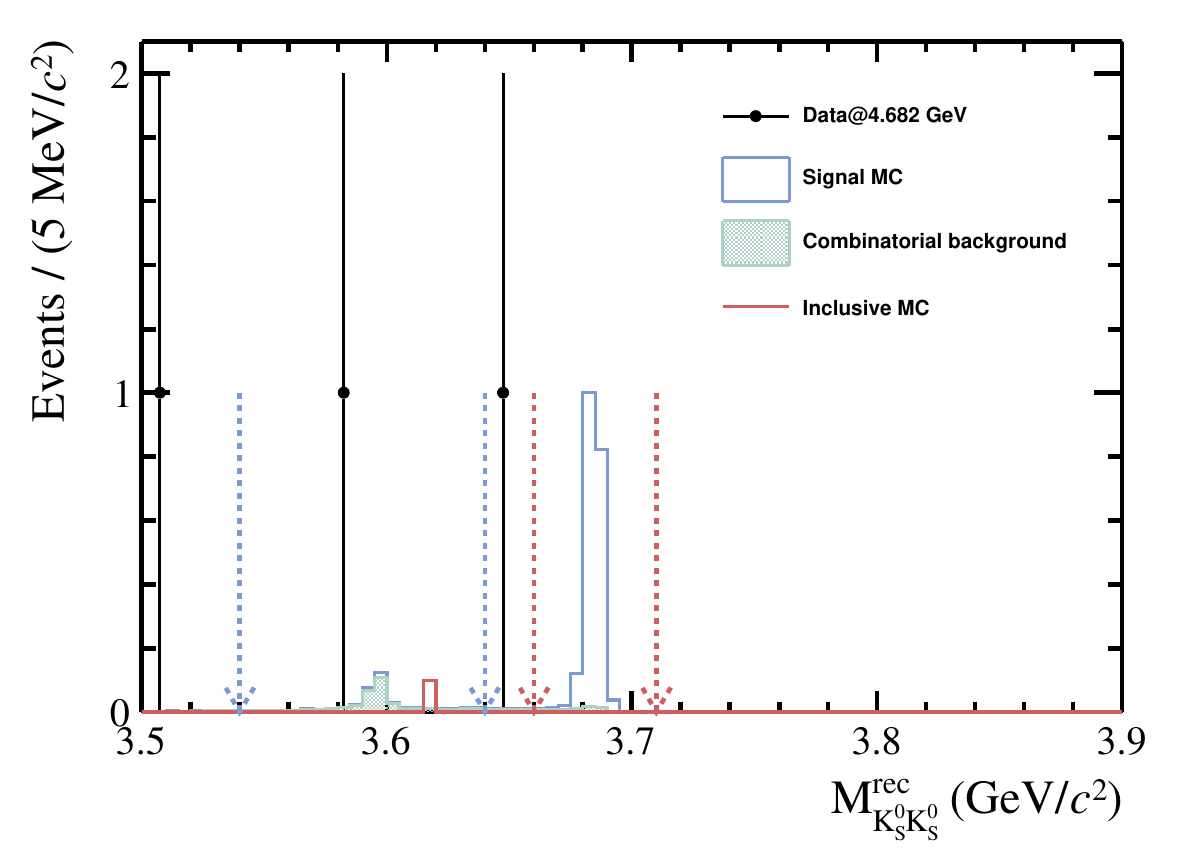}
\includegraphics[width=0.45\textwidth]{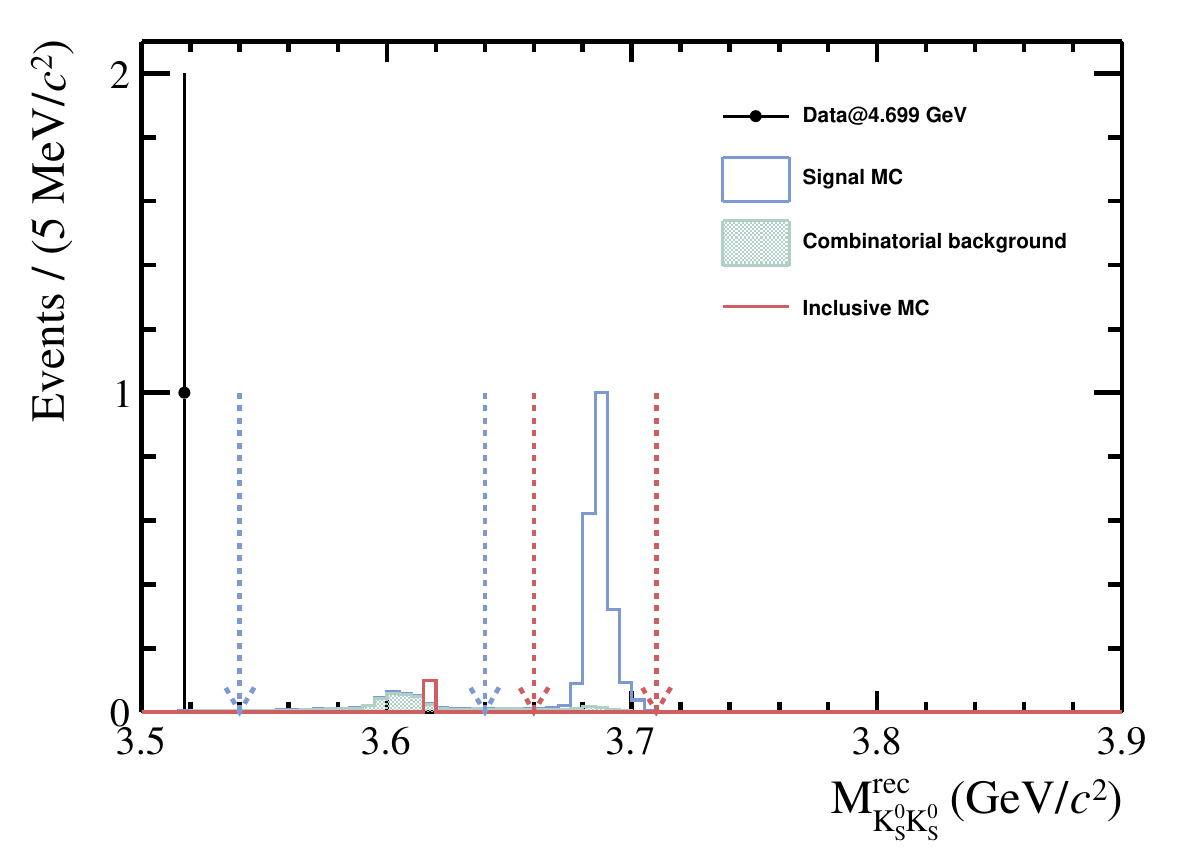}
\includegraphics[width=0.45\textwidth]{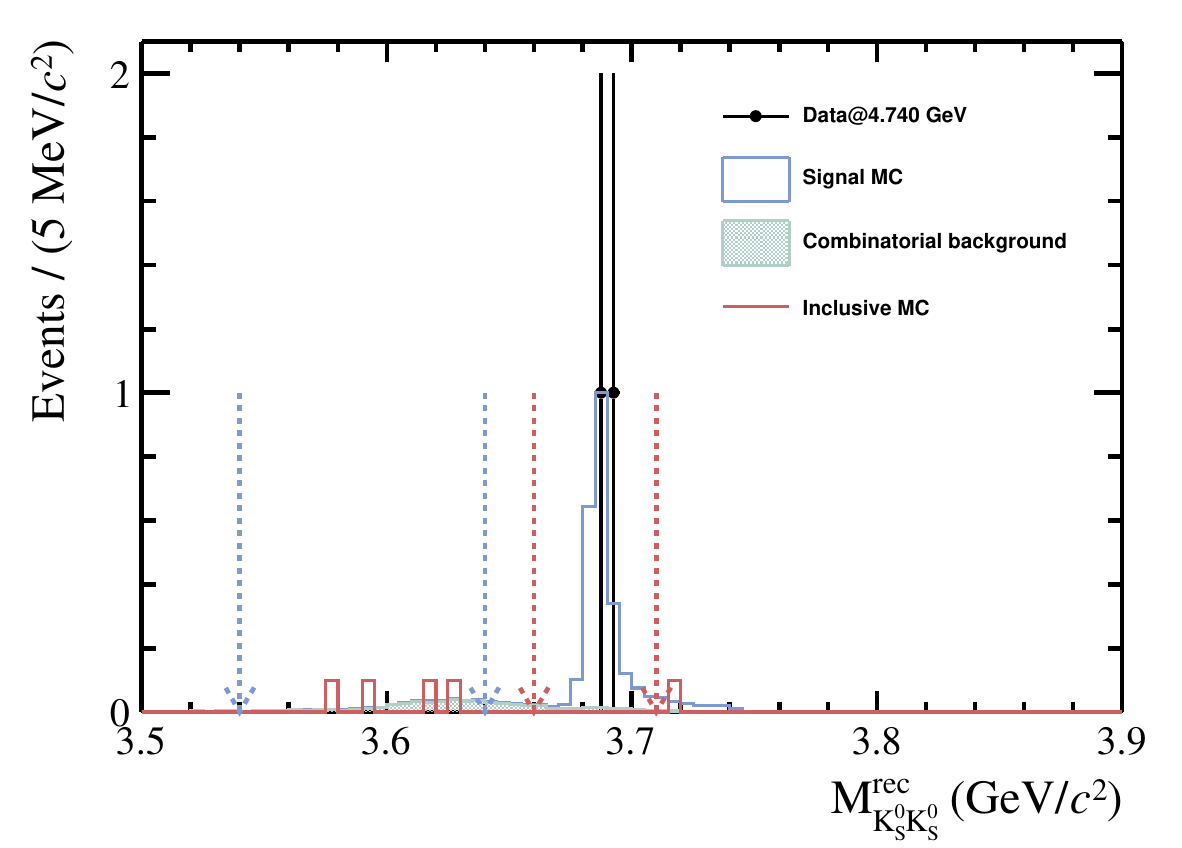}
\includegraphics[width=0.45\textwidth]{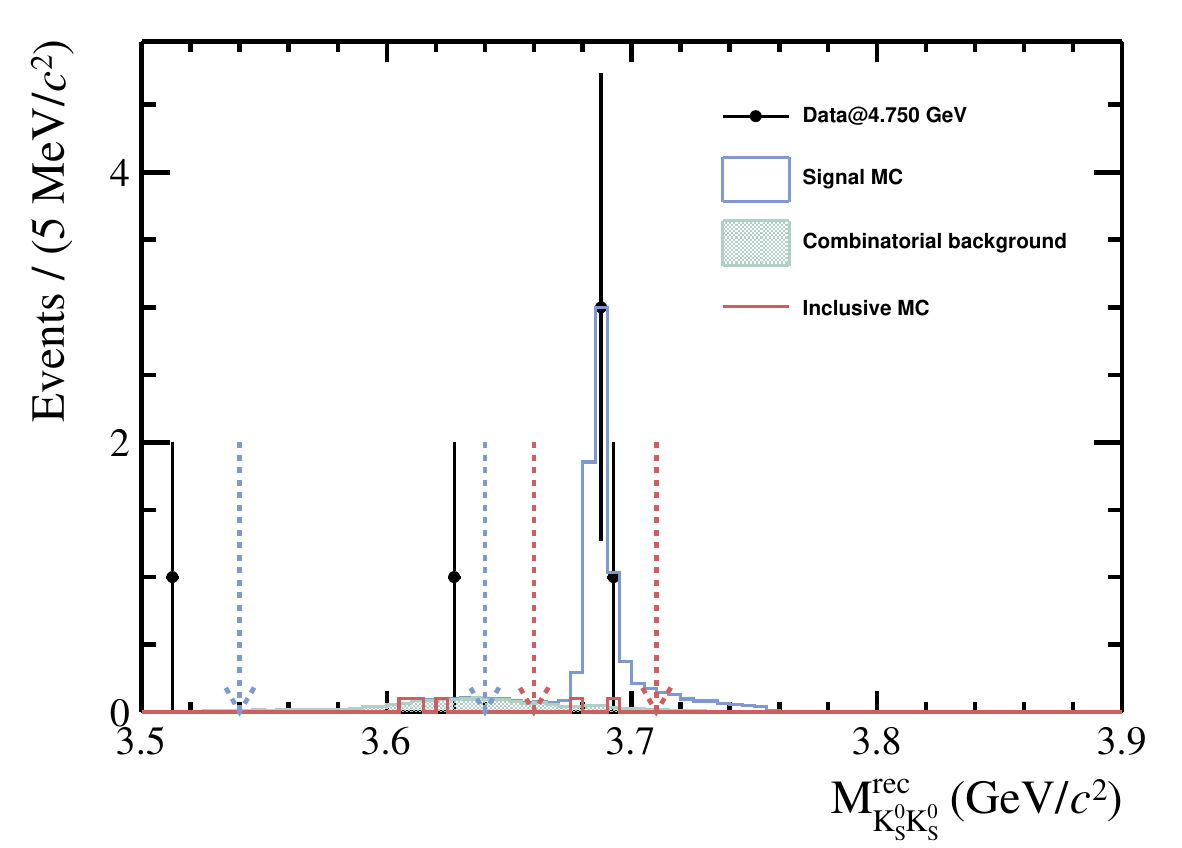}
\includegraphics[width=0.45\textwidth]{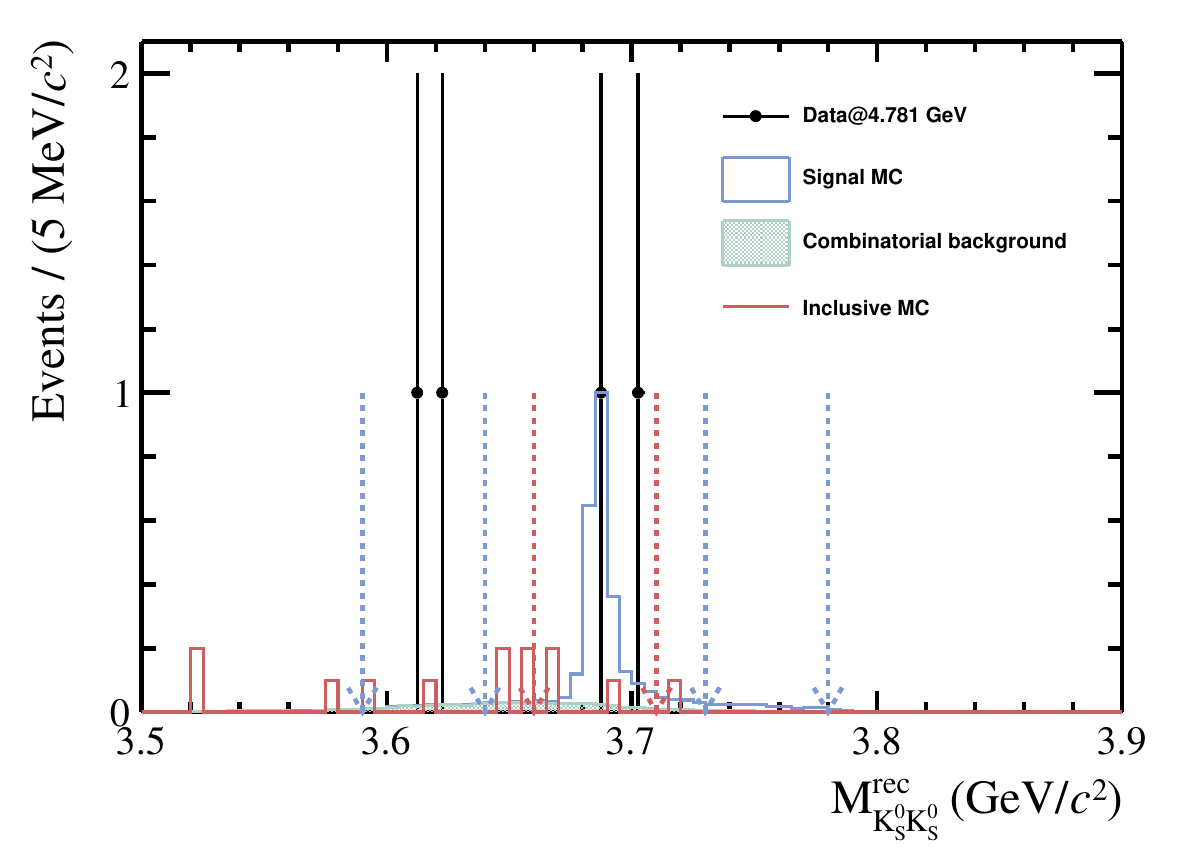}
\includegraphics[width=0.45\textwidth]{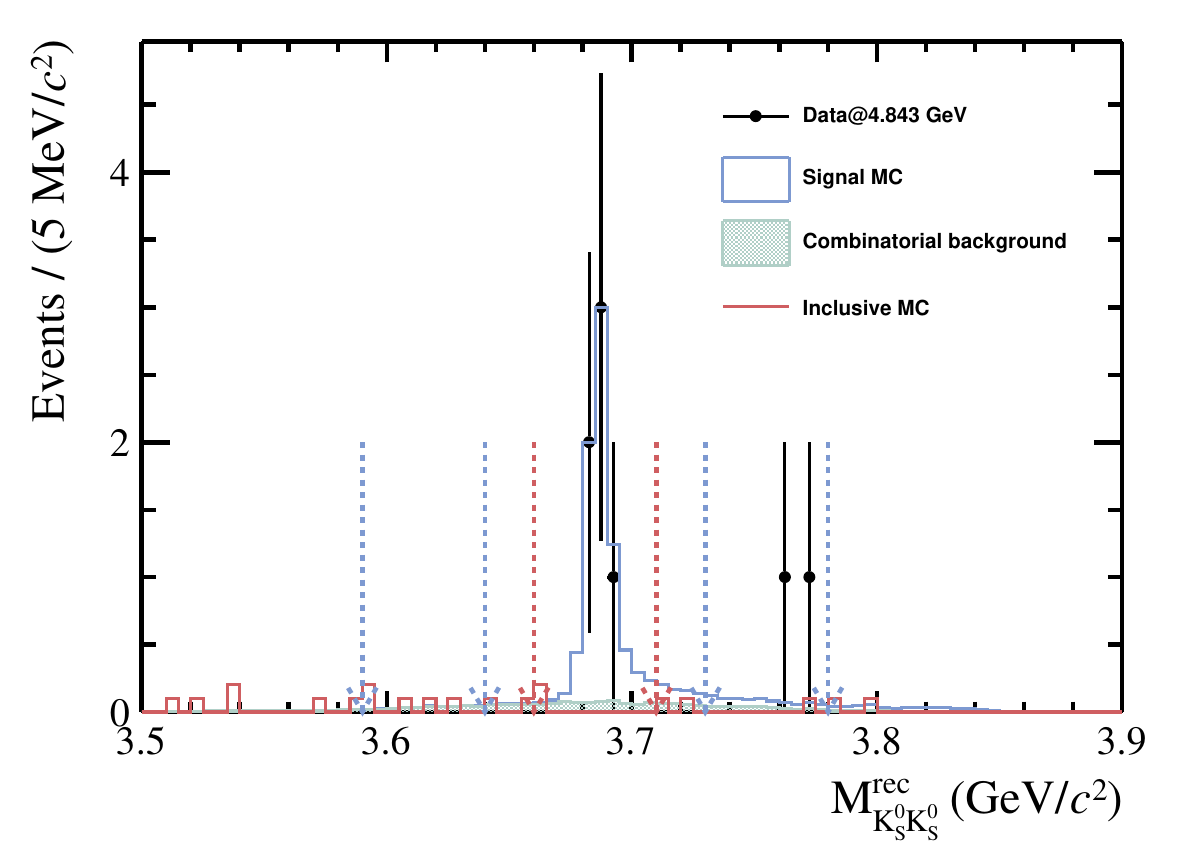}
\includegraphics[width=0.45\textwidth]{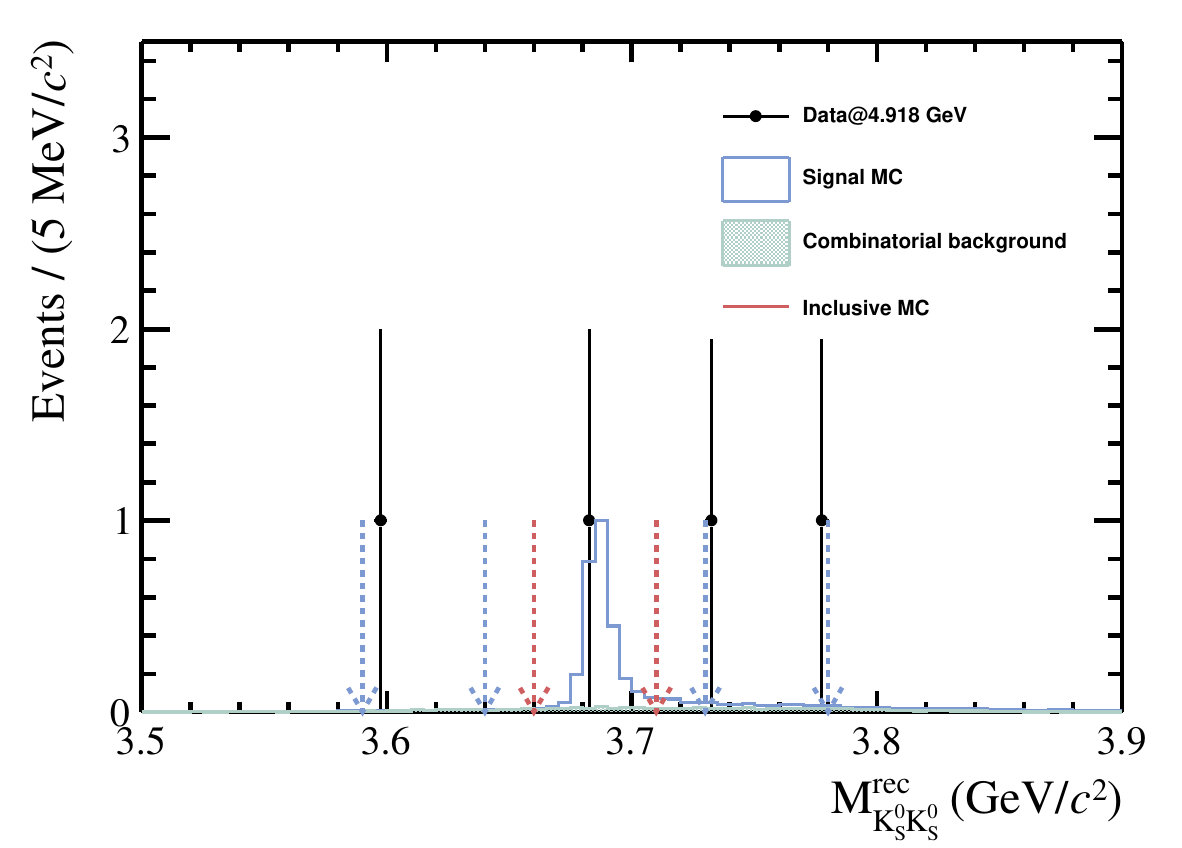}
\includegraphics[width=0.45\textwidth]{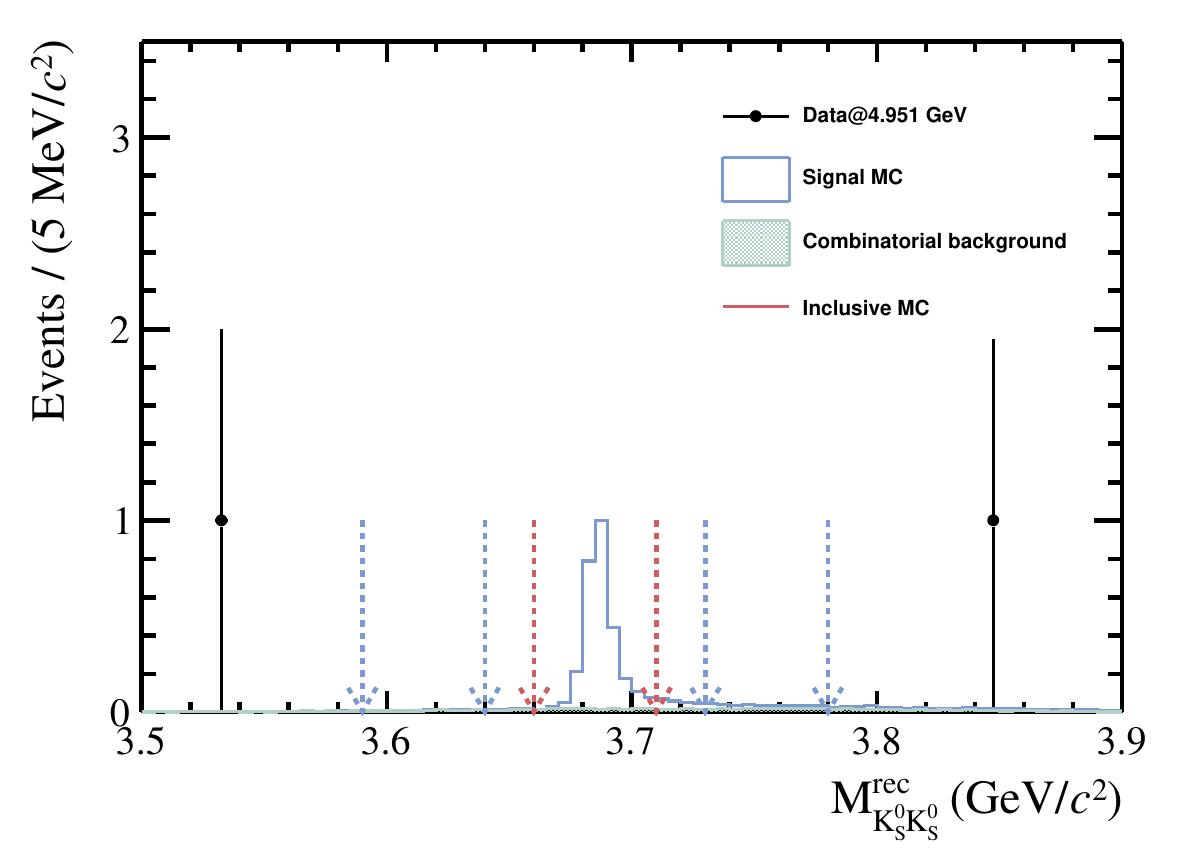}
\end{center}
\caption{
The $M_{K^0_S K^0_S}^{\rm rec}$ distributions for each c.m.\ energy. The black dots with error bars are data, and the red histograms are background contributions estimated by the inclusive MC sample, normalized according to the corresponding integrated luminosity. The blue histograms are the signal MC samples normalized according to the maximum number of events in any bin. The green histograms are the background from multiple combinations in the signal MC sample. The regions within the red and blue arrows indicate the $\psi(3686)$ signal and sideband regions, respectively.}
\label{fig:all psip}
\end{figure*}

\begin{figure*}[htbp]
\begin{center}
\includegraphics[width=0.45\textwidth]{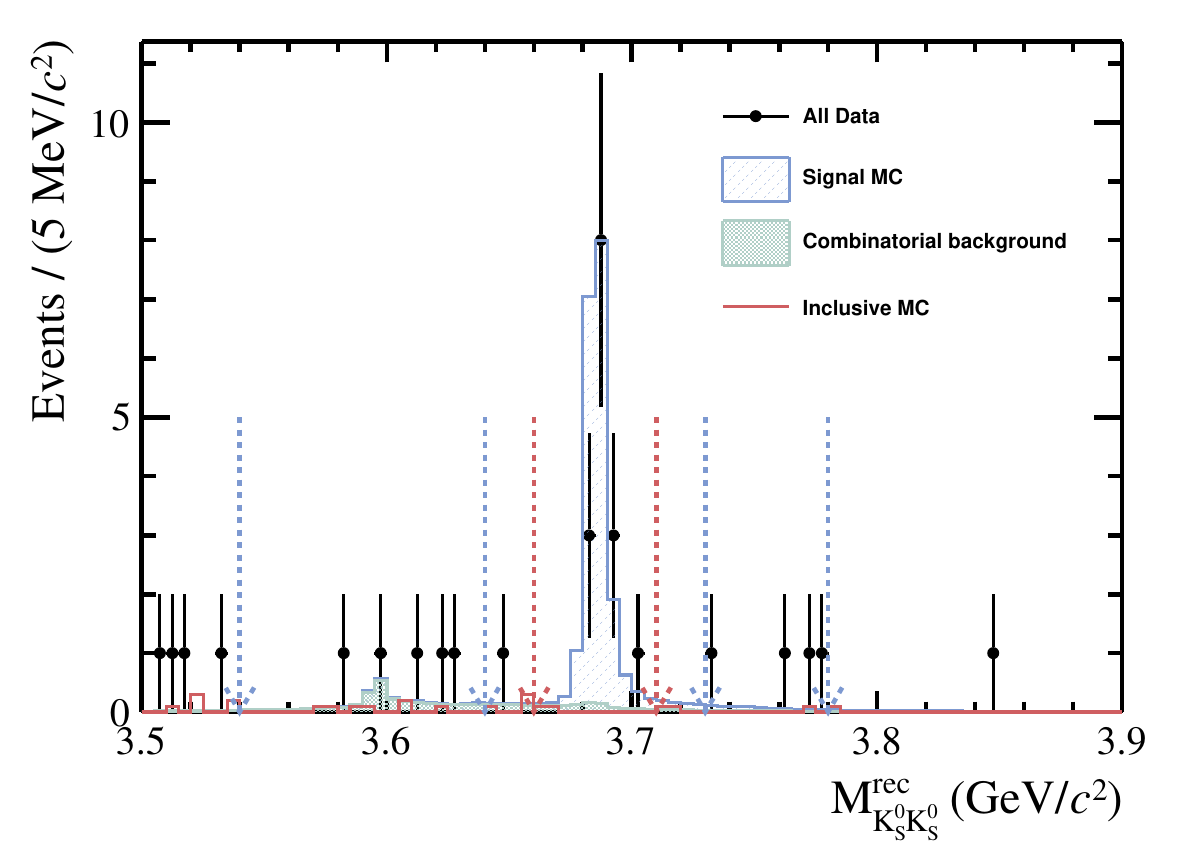}
\put(-90,-8){(a)}
\includegraphics[width=0.45\textwidth]{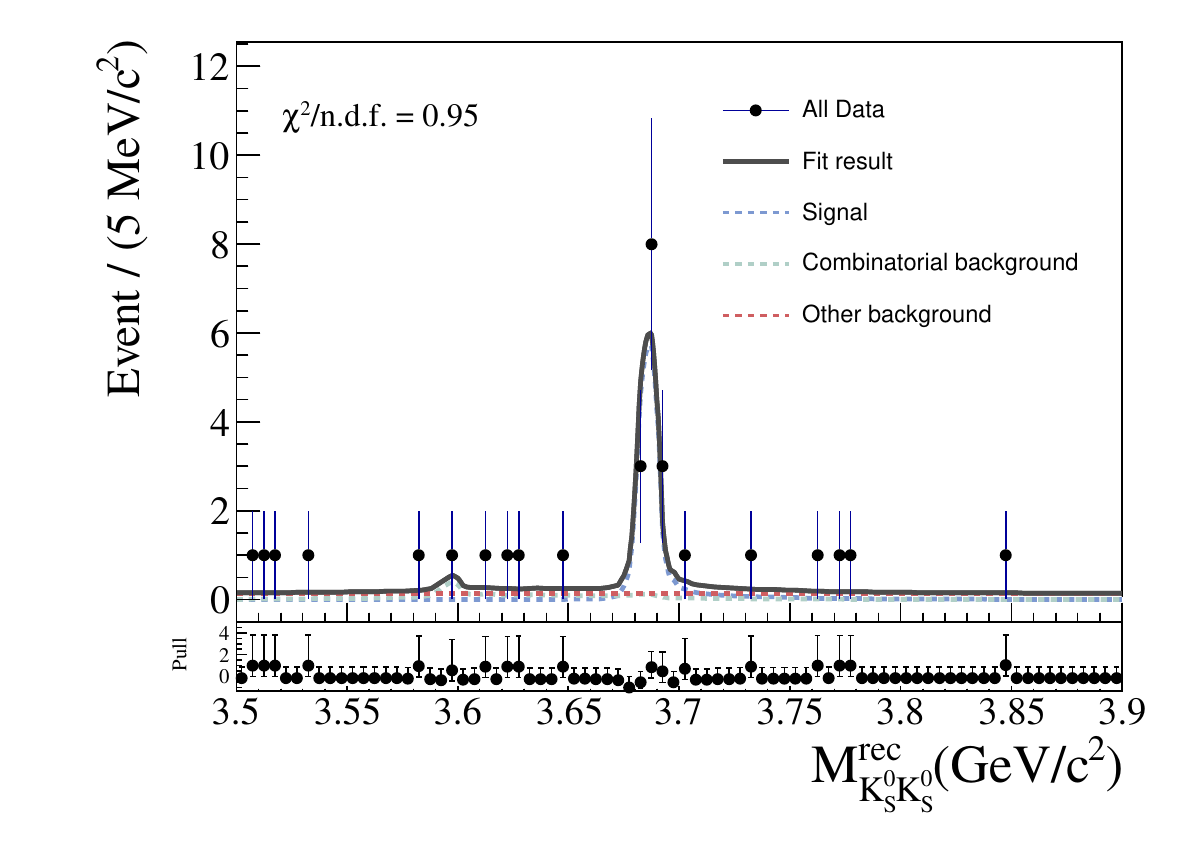}
\put(-90,-8){(b)}
\end{center}
\caption{ (a)~The $M_{K^0_S K^0_S}^{\rm rec}$ distributions from all c.m.\ energies combined. 
The black dots with error bars are data, and the red histograms are background contributions estimated by the inclusive MC sample, normalized according to the corresponding integrated luminosity. The blue histograms are the signal MC samples normalized according to integrated luminosity. The green histograms are the background from multiple combinations in the signal MC sample. The regions within the red and blue arrows indicate the $\psi(3686)$ signal and sideband regions, respectively.
(b)~The fit to the $M_{K^0_S K^0_S}^{\rm rec}$ distributions shown in (a). The black dots with error bars are data, and the solid black line represents the fitting result. The blue dashed curves are the signal components, the green dashed curves are the multi-combination background components, and the red dashed curves are the other background components. 
}
\label{fig:Allcs}
\end{figure*}

The fraction of events containing more than one combination is $8.9\%$, as estimated by the signal MC simulation at $\sqrt{s}=4.843\,\gev$, but it is 0 events in data. The fractions in other samples are at a similar level. The background contribution from multiple combinations in data is negligible. For events with multiple survived combinations in the signal MC simulation, the best one is identified using $\chi^2_{\rm match}=\sum_{i=1}^{4}(\textbf{p}_{i}^{\rm rec}-\textbf{p}_{i}^{\rm truth})^{2}$,
where $\textbf{p}_{i}^{\rm rec}$ is the 4-momenta of the reconstructed pion, and $\textbf{p}_{i}^{\rm truth}$ is from the MC truth information. The $\ks\ks$ recoil mass spectrum ($M_{K^0_S K^0_S}^{\rm rec}$) from wrong combinations exhibits a flat distribution.

The inclusive MC sample is used to study remaining background contributions. The background is found to be negligible, and smoothly distributed in the $M_{K^0_S K^0_S}^{\rm rec}$ distribution, as shown in figure~\ref{fig:all psip}.

\section{Cross section and confidence interval}

Figures~\ref{fig:all psip} and~\ref{fig:Allcs} show the $M_{K^0_S K^0_S}^{\rm rec}$ distributions for accepted candidates at each c.m.\ energy point individually and combined, respectively. The signal yield is obtained by counting events in the $\psip$ signal region (3.66,~3.71)\,$\gev/c^2$. The number of background events in the signal region is estimated using the sideband region, defined as (3.54,~3.64)\,$\gev/c^2$ when $\sqrt{s}<4.77\,\gev$ and $(3.59,~3.64)\cup (3.73,~3.78)\,\gev/c^2$ when $\sqrt{s}>4.77\,\gev$. Therefore, the normalized ratio $f$ of sideband to signal regions is 0.50. The number of expected background events $N_{\rm b}$ is estimated with $f \cdot N_{\rm sdb}$, where $N_{\rm sdb}$ is the events in the sideband region. The number of observed events $N_{\rm obs}$ consists of both the number of expected background events $N_{\rm b}$ and the number of signal events $N_{\rm sig}$~($N_{\rm sig}=N_{\rm obs}-f \cdot N_{\rm sdb}$) and follows a Poisson distribution.

The Born cross section $\sigma_{\mathrm{Born}}$, is calculated as:
\begin{equation}\label{func:dressed cs}
\begin{split}
\sigma_{\mathrm{Born}}&=\frac{N_{\rm{sig}}}
{\mathcal{L}\cdot \epsilon_{r}\cdot(1+\delta)\cdot\delta^{\rm {VP}}},
\end{split}
\end{equation}
where $\epsilon_{r} = {\cal{B}}^{2}(K^0_S \to\pi^+\pi^-)\cdot{\cal{B}}(\psip\to J/\psi X)\cdot \epsilon\cdot {\cal{B}}(J/\psi\to ll)$, $N_{\rm sig}$ is the number of $\ks\ks\psip$ signal events, ${\cal{L}}$ is the integrated luminosity, ${\cal{B}}$ is the branching fraction for each decay, $\epsilon$ is the detection efficiency obtained by subtracting the normalized efficiency in the sideband region from the efficiency in the signal region, ($1+\delta$) is the ISR correction factor, and $\delta^{\rm {VP}}$ is the vacuum polarization factor  taken from ref.~\cite{QED}. Since there is no significant signal, the cross section line-shape from $e^+e^-\to K^+ K^-\psip$~\cite{BESIII-kkpsip} is used for the calculation of the detection efficiency and ISR factor by the method described in ref.~\cite{wenyu_liangliang_iter}. The obtained results are summarized in table~\ref{tab:result}.

The confidence interval (C.I.) of the cross section ($\sigma_{\rm{Born}}^{\rm{C.I.}}$) is determined by replacing $N_{\rm{sig}}$ with its C.I.\ $N_{\rm{sig}}^{\rm{C.I.}}$. The value of $N_{\rm{sig}}^{\rm{C.I.}}$ is determined by counting the number of events in the $\psi(3686)$ signal and sideband regions and using a frequentist method~\cite{Rolke:2004mj} with an unbounded profile likelihood. Assuming the signal and background yields follow a Poisson distribution and the detection efficiency follows a Gaussian distribution, {\textsc{trolke}} package~\cite{TROLK} in the {\textsc{cernroot}} framework~\cite{root} is used to determine the C.I. of the cross section. The obtained results are also summarized in table~\ref{tab:result}.

The significance of the signal process at each c.m.\ energy is calculated by $\cal{P}$-value~\cite{Cowan:1998ji}, $Z = \Phi^{-1}(1- {\cal{P}})$, where $Z$ is the significance, and $\Phi^{-1}$ is the quantile of the normal distribution, assuming all observed events are background events and that $N_{\rm obs}$ follows a Poisson distribution. After considering the uncertainty on $N_{\rm b}$~\cite{Cousins:2007yta}, we use  {\textsc{roostats}} package~\cite{RooStats} in the {\textsc{cernroot}} framework~\cite{root} to obtain the statistical significance at each c.m.\ energy.

The significance of the signal process from all combined data samples is estimated by fitting to the total $M_{K^0_S K^0_S}^{\rm rec}$ distribution shown in figure~\ref{fig:Allcs}(b). In the fit, signal events are described by the total signal MC shape, which is combined across different energy points by normalizing to the luminosity and detection efficiency. The multiple combinations are excluded from the signal MC shape by matching to the generated MC information. The contribution from multiple combinations in signal process is modeled using the corresponding MC shape, extracted from signal MC process with multiple combinations only. Other background contributions are described using a constant function. The signal yield and the number of other background events are free parameters, while the ratio of the number of multi-combinatorial background events and the number of signal events is fixed to that determined in the signal MC sample. The change in log likelihood values ($\Delta(-2\rm{ln}\cal{L})$), found by simultaneously removing the signal shape and the multi-combination background shape from the fit, is used to calculate the significance. The effects of the fitting range and background shape are considered, and in all cases, the statistical significance is above 6.0$\sigma$.

\begin{figure*}[htbp]
\begin{center}
\includegraphics[width=0.45\textwidth]{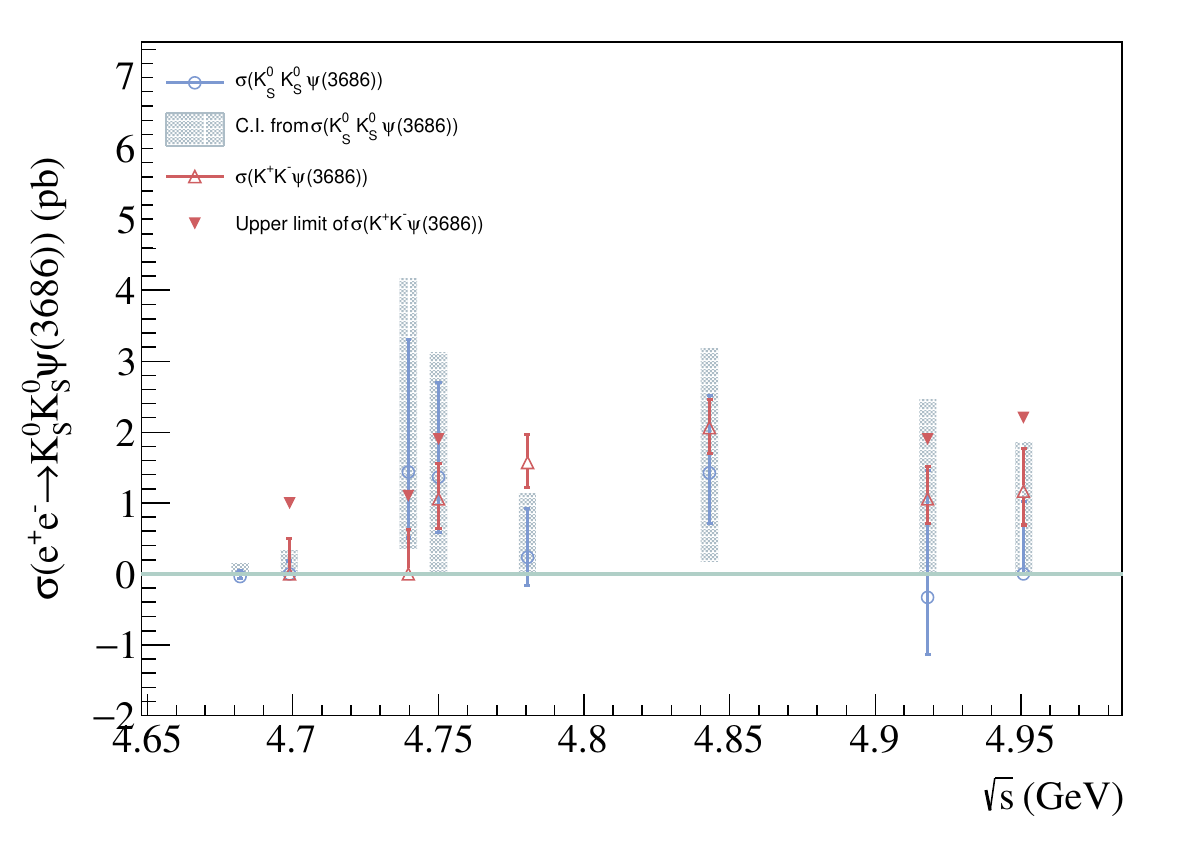}
\put(-90,-8){(a)}
\includegraphics[width=0.45\textwidth]{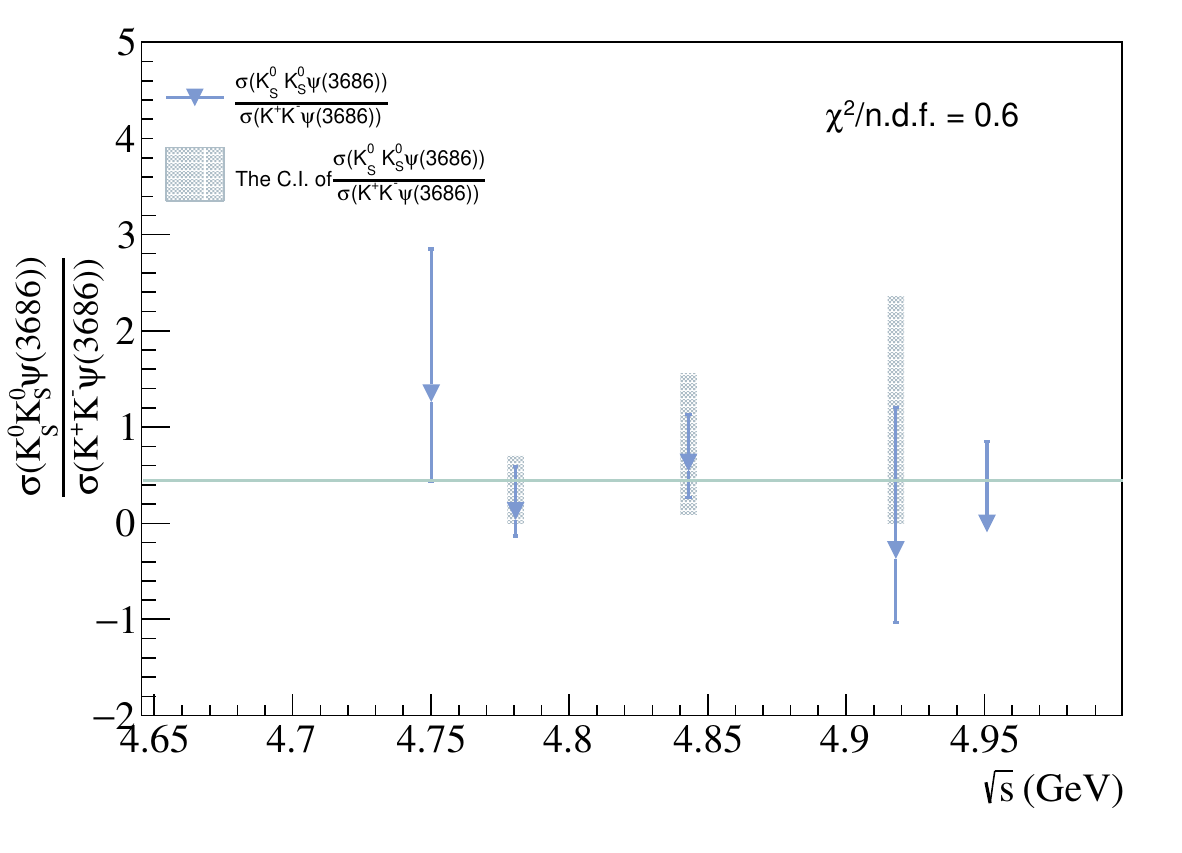}
\put(-90,-8){(b)}
\end{center}
\caption{ (a) The Born cross section and C.I. of $\sigma(e^+e^-\to K^0_S K^0_S \psip)$, where the blue dots with error bars are the measured Born cross sections, and the gray bands are the corresponding C.I. The Born cross section and the upper limit of the 90\% confidence interval of $K^+K^-\psi(3686)$ are also shown in red, based on the published results~\cite{BESIII-kkpsip}. The solid green line represents the cross section at 0. (b) The ratio $\frac{\sigma(e^+e^-\to K^0_S K^0_S \psip)}{\sigma(e^+e^-\to K^+ K^- \psip)}$ at each data sample, where the gray bands are the corresponding C.I. The error bars include both statistical and systematic uncertainties, and the solid curve is the fit result.
}
\label{fig:cs}
\end{figure*}

Figure~\ref{fig:cs}(a) shows the Born cross section and the C.I.\ of $e^+e^-\to K^0_S K^0_S \psip$ at each c.m.\ energy. The ratio of $\sigma(e^+e^-\to K^0_S K^0_S \psip)/\sigma(e^+e^-\to K^+ K^- \psip)$ is shown in figure~\ref{fig:cs}(b), where the cross sections of $e^+e^-\to K^+ K^-\psip$ are quoted from the measurements at BESIII~\cite{BESIII-kkpsip}. The error bars shown in this figure include both statistical and systematic uncertainties of the cross section measurements. The common sources of systematic uncertainties, such as the luminosity, the ISR correction factor, and the $J/\psi$ mass window, cancel between the two measurements. At energies 4.682, 4.699, and 4.740\,$\gev$, the ratios are zero, and are not included in the fitting process. Fitting the ratios with a constant function leads to $\mathcal{R}=\sigma(e^+e^-\to K^0_S K^0_S \psip)/\sigma(e^+e^-\to K^+ K^- \psip) = 0.45 \pm 0.25$. The result is consistent with the expectation from isospin symmetry~\cite{Bjorken:1965sts}. At 4.781, 4.843, and 4.918\,$\gev$, the C.I. of the cross section ratios obtained by dividing the C.I. of $K^0_SK^0_S\psi(3686)$, by the cross section of $K^+K^-\psi(3686)$, are (0,~0.70), (0.10,~1.46), and (0,~2.36), respectively. These ratios include both statistical and systematic uncertainties, with the same systematic uncertainty canceling. The uncertainty of $K^+K^-\psi(3686)$ is taken into account when calculating the C.I. of $K^0_SK^0_S\psi(3686)$ using {\textsc{trolke}} package~\cite{TROLK}. The results are shown in figure~\ref{fig:cs}(b). The C.I. at 4.750 and 4.951\,$\gev$ are not included due to the large uncertainty of the cross section of $K^{+}K^{-}\psi(3686)$.

\begin{figure*}[htbp]
\begin{center}
\includegraphics[width=0.45\textwidth]{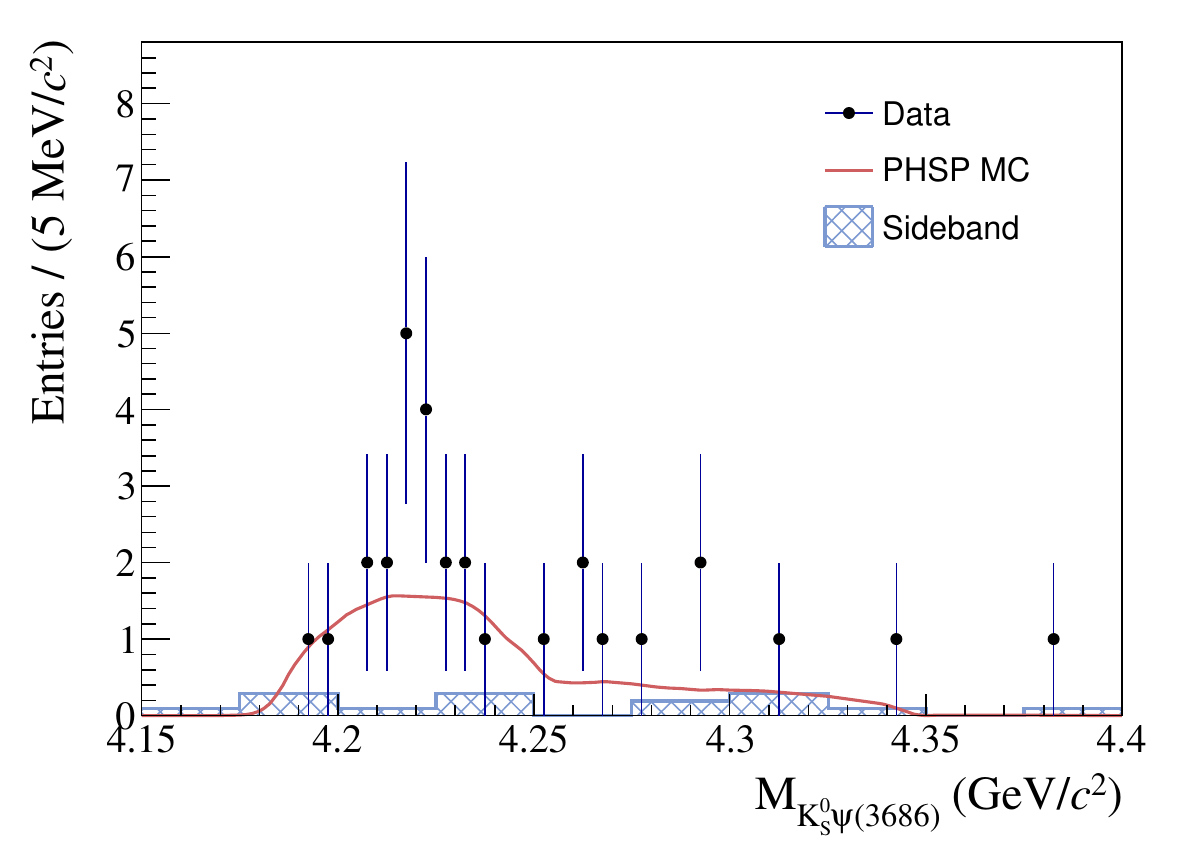}
\put(-90,-8){(a)}
\includegraphics[width=0.45\textwidth]{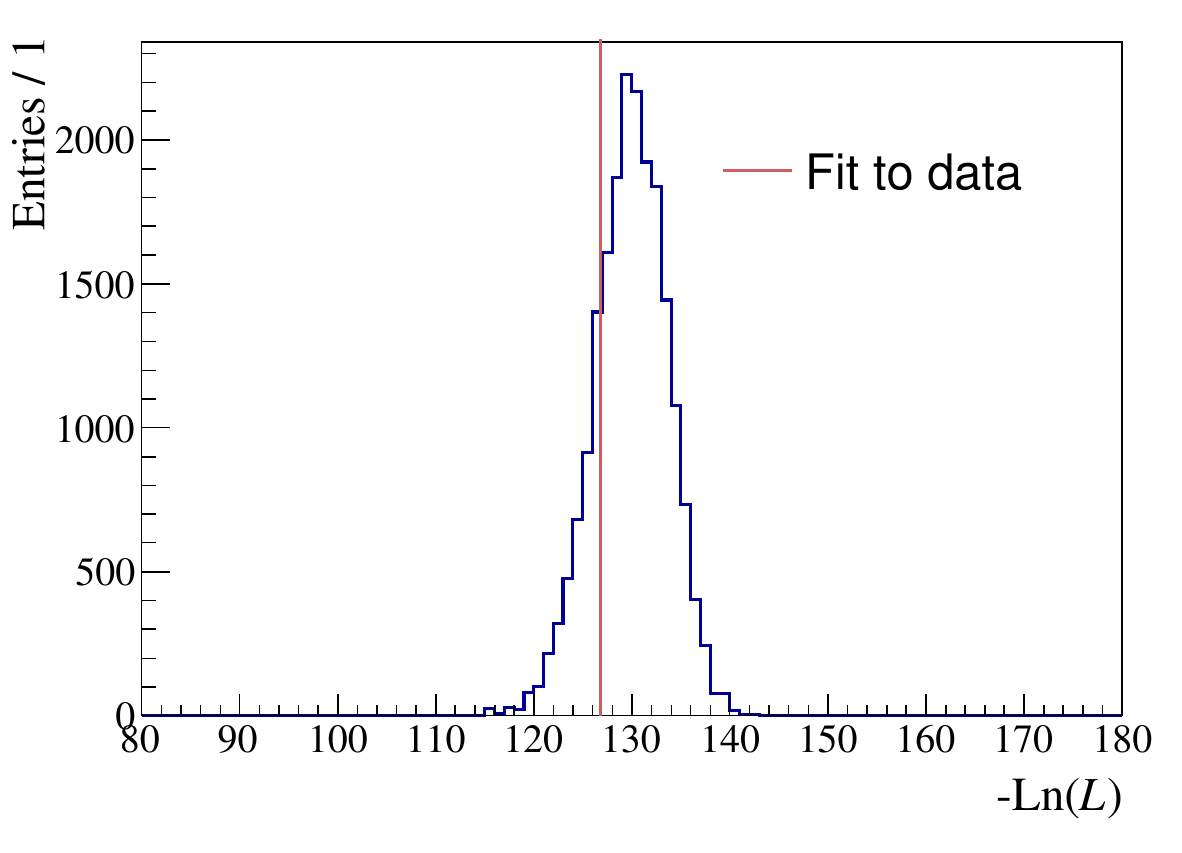}
\put(-90,-8){(b)}
\end{center}
\caption{ (a) The $K_{S}^{0}\psip$ invariant mass distribution. The black dots with error bars are data in the $\psip$ signal region at $\sqrt{s}$ from 4.740 to 4.951\,$\gev$, the blue histogram is the $\psi(3686)$ sideband region in data, and the red histogram is the $K_{S}^{0}K_{S}^{0}\psip$ three-body phase space MC sample at $\sqrt{s}$ from 4.740 to 4.951\,$\gev$, normalized according to the cross sections. (b) The log-likelihood distribution obtained from fitting 20,000 sets of toy MC samples, where the blue histogram represents the log-likelihood distribution and the red vertical line indicates the log-likelihood value for the fit to the data.
}
\label{fig:all psip Zcs}
\end{figure*}

Intermediate states in the $K^0_S\psip$ system are investigated using data samples with $\sqrt{s}$ from 4.740 to 4.951\,$\gev$. As each event contains two $K_{S}^{0}$ mesons, we combine both $K_{S}^{0}$ mesons with the $\psi(3686)$ for the invariant mass distribution of $K_{S}^{0}\psi(3686)$ (figure~\ref{fig:all psip Zcs}), resulting in each event appearing twice in the distribution. Assuming the observed events are all from $K_{S}^{0}K_{S}^{0}\psip$ three-body phase space, the $K_{S}^{0}\psip$ invariant mass distribution from MC simulation is displayed in the same figure. To evaluate the significance of possible contributions in addition to the $K_{S}^{0}K_{S}^{0}\psip$ three-body phase space, 20,000 sets of toy MC samples are produced using the sum of the phase space MC shape and the non-$\psip$ events estimated from the sideband region in the data. In each toy MC sample, the number of entries is set to match the entries in data, as shown in figure~\ref{fig:all psip Zcs}(a). Fitting the toy MC samples with the same PDF as used in producing these samples, the $-\ln L$ distribution is shown in figure~\ref{fig:all psip Zcs}(b). The red vertical line indicates the $-\ln L$ value obtained from the fit to the data. The $\cal{P}$-value determined from the toy MC samples is 0.197, corresponding to a significance of $0.8\sigma$ for the deviation between the data and the model with the phase space contribution only.

\section{Systematic uncertainty}

The systematic uncertainties in the cross section measurement are caused by the luminosity, the tracking, the $K^0_S$ reconstruction, the branching fraction, the kinematic fit, the ISR correction factor, the $J/\psi$ mass window, and the $\psi(3686)$ signal and sideband regions. The systematic uncertainties are summarized in table~\ref{tab:sys total} where the total systematic uncertainty is calculated as a sum in quadrature of all sources of uncertainty assuming they are independent.

The systematic uncertainties in the determination of the C.I. of the cross section can be classified as either additive or multiplicative terms. The additive terms include the $\psi(3686)$ signal and sideband region selections where these regions are varied, and the largest C.I. is chosen. The multiplicative systematic uncertainties are considered in the calculations of C.I.s by using the {\textsc{trolke}} package.

The integrated luminosity is measured by Bhabha events, with an uncertainty of 0.6\%~\cite{cms-lumi-round1314}. The systematic uncertainty from tracking is 1.0\% per track~\cite{Zc3900-1}. The systematic uncertainty associated with $K^0_S$ reconstruction is studied with the control sample $\jpsi\to K^{*}(892)^{\pm}K^{\mp}$~\cite{ref:ks}, and is assigned to be 1.2\% per $K^0_S$. The uncertainties on the quoted branching fractions of different decays are taken from the PDG~\cite{PDG2018}. The helix parameters of the charged tracks are corrected in simulation to improve the agreement of the $\chi^2$ from the kinematic fit between data and MC simulation~\cite{BESIII:2012mpj}. The systematic uncertainty from the kinematic fit is assigned as the difference in the efficiencies with and without this correction. The uncertainty from the ISR correction is estimated by sampling the parameters of the cross section line-shape $K^+K^-\psip$ 1000 times according to the covariance matrix given by the fit to the cross section~\cite{BESIII-kkpsip}. The resultant distribution of $(1+\delta)\cdot\epsilon$ is fitted with a Gaussian function and the standard deviation is assigned as the systematic uncertainty. The 1.0\% is accounted as a systematic uncertainty on the $J/\psi$ mass window according to ref.~\cite{intro-BESIII-ksksjpsi}.

In the cross section measurement, the systematic uncertainty due to the choices of $\psip$ signal and sideband regions are estimated by varying the boundaries of these sideband regions. To avoid the difference caused by statistical fluctuations, 1000 sets of toy MC samples are generated based on the fit results, then we fit the distribution of $M_{K^0_S K^0_S}^{\rm rec}$. In the fit, the signal is described by the MC shape, and the background is described by a first-order Chebyshev function. Then, $[(N/\epsilon)_{\rm nominal}-(N/\epsilon)_{\rm change}]/(N/\epsilon)_{\rm nominal}$ is calculated and modeled with a Gaussian function. The mean value of the Gaussian function represents the difference between the modified and nominal regions, and is taken as the corresponding uncertainty.

\begin{table*}[htbp]
\caption{ Relative systematic uncertainties (\%) in the cross section measurement at each c.m.\ energy point, where the sources marked with * indicate additive uncertainties and the rest are multiplicative uncertainties. }
\label{tab:sys total}
\begin{center}
\begin{tabular}{ccccccccc}
\hline
Source/$\sqrt s$~(GeV)  &4.682 &4.699 &4.740 &4.750 &4.781 &4.843 &4.918 &4.951 \\
\hline
Luminosity &0.6 &0.6&0.6&0.6&0.6&0.6&0.6&0.6\\
Tracking &2.0 &2.0 &2.0 &2.0 &2.0 &2.0 &2.0 &2.0 \\
$K^0_S$ reconstruction &2.4 &2.4 &2.4 &2.4 &2.4 &2.4 &2.4 &2.4 \\
Branching fraction &1.1 &1.1 &1.1 &1.1 &1.1 &1.1 &1.1 &1.1\\
Kinematic fit &0.2 &0.3 &0.3 &0.3 &0.2 &0.2 &0.2 &0.2\\
ISR correction &1.0 &1.2 &3.0 &1.5 &0.6 &0.2 &0.6 &0.5\\
$J/\psi$ mass window &1.0 &1.0 &1.0 &1.0 &1.0 &1.0 &1.0 &1.0\\
$\psip$ signal region* &2.8 &2.8 &2.8 &2.8 &2.8 &2.8 &2.8 &2.8\\
$\psip$ sideband region* &4.8 &4.8 &4.8 &4.8 &4.8 &4.8 &4.8 &4.8\\
\hline
Sum &6.7 &6.7 &7.2 &6.7 &7.3 &6.6 &6.6 &6.6 \\
\hline
\end{tabular}
\end{center}
\end{table*}

\section{Summary} 

In summary, the $e^+e^-\to K^0_S K^0_S\psip$ process is studied using data samples accumulated at c.m.\ energies from 4.682 to 4.951\,$\gev$ with the BESIII detector at BEPCII. The cross sections of $e^+e^-\to K^0_S K^0_S\psip$ at different c.m.\ energy points are measured for the first time. The statistical significance of $e^+e^-\to K^0_S K^0_S\psip$ is $6.3\sigma$ when summing over all data samples. The C.I.\ at the 90\% confidence level is provided for each data sample. In addition, the ratio $\sigma(e^+e^-\to K^0_S K^0_S \psip)/\sigma(e^+e^-\to K^+ K^- \psip)$ is determined to be $0.45 \pm 0.25$, where the uncertainty includes both statistical and systematic uncertainties. The results agree with the predictions based on isospin symmetry within uncertainties. We search for the $Z_{cs}$ state in the $Z_{cs}\to K^0_S \psip$ decay using data samples taken at $\sqrt{s}=4.740-4.951\,\gev$. The $K^0_S\psi(3686)$ invariant mass distribution is  consistent with three-body phase space, and no obvious structure is found. Larger data samples are needed to further investigate the vector states and the $Z_{cs}$ in this process.

\textbf{Acknowledgement}

The BESIII Collaboration thanks the staff of BEPCII and the IHEP computing center for their strong support. This work is supported in part by National Key R\&D Program of China under Contracts Nos. 2020YFA0406300, 2020YFA0406400, 2023YFA1606000; National Natural Science Foundation of China (NSFC) under Contracts Nos. 12375070, 11635010, 11735014, 11935015, 11935016, 11935018, 12025502, 12035009, 12035013, 12061131003, 12192260, 12192261, 12192262, 12192263, 12192264, 12192265, 12221005, 12225509, 12235017, 12361141819; the Chinese Academy of Sciences (CAS) Large-Scale Scientific Facility Program; the CAS Center for Excellence in Particle Physics (CCEPP); Joint Large-Scale Scientific Facility Funds of the NSFC and CAS under Contract No. U2032108, U1832207; Shanghai Leading Talent Program of Eastern Talent Plan under Contract No. JLH5913002; 100 Talents Program of CAS; The Institute of Nuclear and Particle Physics (INPAC) and Shanghai Key Laboratory for Particle Physics and Cosmology; German Research Foundation DFG under Contracts Nos. FOR5327, GRK 2149; Istituto Nazionale di Fisica Nucleare, Italy; Knut and Alice Wallenberg Foundation under Contracts Nos. 2021.0174, 2021.0299; Ministry of Development of Turkey under Contract No. DPT2006K-120470; National Research Foundation of Korea under Contract No. NRF-2022R1A2C1092335; National Science and Technology fund of Mongolia; National Science Research and Innovation Fund (NSRF) via the Program Management Unit for Human Resources \& Institutional Development, Research and Innovation of Thailand under Contracts Nos. B16F640076, B50G670107; Polish National Science Centre under Contract No. 2019/35/O/ST2/02907; Swedish Research Council under Contract No. 2019.04595; The Swedish Foundation for International Cooperation in Research and Higher Education under Contract No. CH2018-7756; U. S. Department of Energy under Contract No. DE-FG02-05ER41374



\clearpage

\section*{The BESIII collaboration}
\addcontentsline{toc}{section}{The BESIII collaboration}
\begin{small}
M.~Ablikim$^{1}$, M.~N.~Achasov$^{4,c}$, P.~Adlarson$^{76}$, O.~Afedulidis$^{3}$, X.~C.~Ai$^{81}$, R.~Aliberti$^{35}$, A.~Amoroso$^{75A,75C}$, Q.~An$^{72,58,a}$, Y.~Bai$^{57}$, O.~Bakina$^{36}$
% ...
\\
\vspace{0.2cm}
(BESIII Collaboration)\\
{\it
$^{1}$ Institute of High Energy Physics, Beijing 100049, People's Republic of China\\
% ...
$^{3}$ Bochum  Ruhr-University, D-44780 Bochum, Germany\\
$^{4}$ Budker Institute of Nuclear Physics SB RAS (BINP), Novosibirsk 630090, Russia\\
% ...
$^{35}$ Johannes Gutenberg University of Mainz, Johann-Joachim-Becher-Weg 45, D-55099 Mainz, Germany\\
$^{36}$ Joint Institute for Nuclear Research, 141980 Dubna, Moscow region, Russia\\
% ...
$^{57}$ Southeast University, Nanjing 211100, People's Republic of China\\
$^{58}$ State Key Laboratory of Particle Detection and Electronics, Beijing 100049, Hefei 230026, People's Republic of China\\
% ...
$^{72}$ University of Science and Technology of China, Hefei 230026, People's Republic of China\\
% ...
$^{75}$ University of Turin and INFN, (A)University of Turin, I-10125, Turin, Italy; (B)University of Eastern Piedmont, I-15121, Alessandria, Italy; (C)INFN, I-10125, Turin, Italy\\
$^{76}$ Uppsala University, Box 516, SE-75120 Uppsala, Sweden\\
% ...
$^{81}$ Zhengzhou University, Zhengzhou 450001, People's Republic of China\\
$^{a}$ Deceased\\
% ...
$^{c}$ Also at the Novosibirsk State University, Novosibirsk, 630090, Russia\\
% ...
}
\end{small}

\end{document}